\documentclass[fleqn,a4paper,english,10pt]{article}
\usepackage{babel,amsmath,amsfonts,amsgen,amssymb,amsbsy,graphicx,physics,xcolor,flushend,longtable}
\usepackage[numbers,square,sort&compress]{natbib}

\newcommand\rmax{r_0}

\newcommand\Nu{\mathrm{Nu}}

\newcommand\Dh{D_h}
\newcommand\DhNom{\hat{D}_h}

\newcommand\um{u_m}

\newcommand\Pcal{\mathcal{P}}
\newcommand\Scal{\mathcal{S}}

\newcommand\Tb{T_b}
\newcommand\Tw{T_w}
\newcommand\qw{q_w}

\newcommand\Twb{\bar{T}_w}
\newcommand\qwb{\bar{q}_w}

\newcommand\as{a^\ast}
\newcommand\rs{r^\ast}
\newcommand\xs{x^\ast}
\newcommand\ys{y^\ast}
\newcommand\Ts{T^\ast}
\newcommand\us{u^\ast}
\newcommand\lambdas{\lambda^\ast}
\newcommand\nablas{\nabla^\ast}
\newcommand\Ps{\Pcal^\ast}
\newcommand\Ss{\Scal^\ast}
\newcommand\Dhs{D^\ast_h}

\newcommand\Twbs{\bar{T}_w^\ast}
\newcommand\Tbs{T_b^\ast}
\newcommand\hb{\bar{h}}

\newcommand\hide[1]{}

\newcommand{\eqn}[2]{\begin{gather}
#1
\label{#2}
\end{gather}
}
\title{Laminar forced convection characteristics in a round microchannel with shape uncertainty: effect of wall slip}

\author{L.~A.~Sphaier$^{1}$, A. Barletta$^{*,2}$, M. Celli$^{2}$, P. V. Brand\~ao$^{2}$, S. Lazzari$^{3}$, E. Ghedini$^{2}$}
\vspace{12pt}

\date{\small
$^1$ Department of Mechanical Engineering, Universidade Federal Fluminense\\ Rua Passo da Pátria 156, sala 302, bloco D, Niterói, RJ, Brazil\\
$^2$ Department of Industrial Engineering, Alma Mater Studiorum Universit\`a di Bologna\\ Viale Risorgimento 2, 40136 Bologna, Italy\\
$^3$ Department of Architecture and Design, Università degli Studi di Genova\\ Stradone di S.Agostino 37, 16128 Genova, Italy\\
$^*$ Email: antonio.barletta@unibo.it}

\begin{document}
\maketitle
\begin{abstract}
The shape of a microchannel cross-section is usually affected by a significant uncertainty due to the small hydraulic diameter. Such an uncertainty is indeed present at any scale, but is amplified in smaller scales, becoming significantly important when the hydraulic diameter is smaller than some tenth micrometers. 
In this scenario, this paper is focused on analyzing the sensitivity of the heat and fluid flow characteristics with respect to the channel shape, considering a random modifications in the channel cross section. Forced convection in a fully developed regime with a wall slip is considered, 
and the analysis includes the evaluation of the Fanning friction factor and of the Nusselt number for the H1 thermal boundary condition, considering different slip-flow configurations, as dictated by the slip-length parameter.
The heat and fluid flow problem is solved for one thousand randomly generated channel geometries, based on a circular microchannel which is allowed to have its boundary points move within 10\% of its nominal diameter. The calculated data for $f \Re$ and $\Nu$ are then analyzed, and compared with the nominal values (obtained for smooth channels). The results show that, on an average basis, the roughness effect has a tendency to reduce the Nusselt number while increasing the friction factor, however, with a small number of exceptions.
\end{abstract}

\subsection*{Nomenclature}
\begin{longtable}[l]{ll}
$\langle \cdot\, \rangle$ & mean value\\
$a$ & constant temperature gradient\\
$f$ & Fanning friction factor\\
$k$ & thermal conductivity\\
$\vb n$ & unit normal vector\\
$\Nu$ & Nusselt number\\
$p$ & pressure\\
$\cal P$ & cross-sectional perimeter\\
$\qwb$ & average wall heat flux\\
$r$ & radial coordinate\\
$\rmax$ & duct radius\\
$\Re$ & Reynolds number\\
$\cal S$ & cross-sectional area\\
$T$ & temperature\\
$\Tb$ & bulk temperature\\
$\Twb$ & average wall temperature\\
$u$ & velocity\\
$u_0$ & reference velocity\\
$\um$ & mean flow velocity\\
$x,y,z$ & Cartesian coordinates\\[6pt]
{\em Greek symbols} \\
$\alpha$ & thermal diffusivity\\
$\Delta(\cdot)$ & standard deviation\\
$\lambda$ & slip length \\
$\lambda_T$ & temperature jump parameter \\
$\mu$ & dynamic viscosity\\
$\sigma$ & Poiseuille number \\[6pt]
 {\em Subscripts, superscripts}\hspace{-1.5cm} \\
$(~)^\ast$ & dimensionless quantity\\
$(\bar{~})$ & averaged quantity\\
\end{longtable}
\setcounter{table}{0}

\section{Introduction}

Forced convection in microchannels is of great interest for applications in the field of thermal management \cite{siddiqui2017efficient}. 
An important example of such applications is the regenerative cooling of combustion chamber liners, where there is a large heat flux removal at diverse operation temperatures. 
In addition, studies of microfluidics and microscale heat transfer have a significant impact on the cryogenic cooling employed for the supercomputing chips. 
Also, cooling of concentrator photovoltaics employing microchannels was shown to be an innovative application, as pointed out by \citet{gilmore2018microchannel}.

Miniaturization of heat exchangers for several energy and heat transfer applications often entails the use of microchannels. 
However, due to the small hydraulic diameters involved, a significant uncertainty affects the shape of a microchannel cross-section. 
When the hydraulic diameter becomes smaller than some tenth micrometers, the shape uncertainty induced by the wall roughness, and present at any scale, becomes significantly important. 
%

While there is a broad number of literature studies regarding the effects of slip flow in the analysis of forced convection in microchannels, there seems to be little information regarding the effects of random departures from the average shape of the cross-section. 
Such departures are expected to be important due to the roughness of the solid wall, which depends on the manufacturing techniques and on the surface treatment of the materials. 
Their effects lead to uncertainties in the significant parameters characteristic of the forced convection flow and heat transfer. 
Some authors model the wall roughness via regular geometric patterns, as in the study by \citet{wang2007influence} where the wall roughness is represented through periodic wall functions. 
A similar approach can also be found in \citet{turgay2009effect} and in \citet{croce2007three} where regular distributions of conical peaks have been envisaged. 
Random generation of peaks and valleys of rectangular shape was carried out in the computational study by \citet{croce2005numerical}. 
Experimental data assessing the role of wall roughness were reported by \citet{celata2002experimental} relative to the transition from laminar to turbulent regimes.
Under the previous considerations, this investigation is aimed at performing an analysis of heat and fluid flow in a nominally circular microchannel, allowing for random variations in the cross-section geometry. With this analysis, one can determine how sensitive heat and fluid flow parameters, such as the friction factor and Nusselt number, are to such random variations.

\section{Governing equations}

Consider a straight and nominally round cross-section microchannel and allow for an uncertainty in the actual shape of the cross-section, due to the practical manufacturing characteristics whatever is the adopted fabrication technique. 
Such uncertainty issues may be due to the size of the typical surface roughness relative to the small hydraulic diameter of the microchannel. 
Hence, although the average shape of the microchannel is circular with radius $\rmax$, the perimeter of the cross-section will be denoted $\Pcal$ and the encircled flow area $\Scal$. 
More precisely, $\Pcal$ is a closed boundary which matches a circle within a $10\%$ tolerance, meaning that each point on $\cal P$ has a distance from the centre of the cross-section smaller than $\pm\,\rmax/10$. 
Note that the symbols $\Pcal$ and $\Scal$ will be tacitly employed to denote both the geometric objects and their measures (perimeter and area).
The actual hydraulic diameter of the irregular geometry is defined based on these quantities as:
\eqn{
\Dh \,=\, \frac{4\,\Scal}{\Pcal} ,
}{eq:Dh-def}
while, the nominal hydraulic diameter will be referred to as $\DhNom$, which for an circular duct is simply given by $2\,\rmax$.

Let $(x,y)$ be the Cartesian coordinates on the cross-sectional plane and $z$ be the streamwise coordinate. 
The flow is assumed to be laminar, stationary, incompressible and fully-developed, so that the only nonzero component of the velocity field is that in the $z$ direction, which is denoted with $u$. Due to the local mass balance equation, {\em i.e.} the zero-divergence condition on the velocity, $u$ is independent of $z$, and hence $u=u(x,y)$. Furthermore, the local difference between the pressure and the hydrostatic pressure depends only on $z$ and is denoted as $p(z)$. Under such conditions, the local momentum balance equation is expressed~as
\eqn{
\laplacian{u} = \frac{1}{\mu}\, \dv{p}{z},
}{eq:mom-1}
where $\dd p/\dd z$ is a constant and $\mu$ denotes the dynamic viscosity. Thus, \eqref{eq:mom-1} is a two-dimensional Poisson equation defined in the $(x,y)$ plane. Such an equation is subject to a slip boundary condition
\eqn{
u +\lambda\,\vb{n} \cdot \,\nabla  u= 0 \qc (x,y) \in \Pcal.
}{2}
where $\lambda > 0$ is  the slip length \cite{bruus2007theoretical} and the two-dimensional vector $\vb{n}$ is the unit outward normal to $\cal P$.

The local energy balance is given as:
\eqn{
\alpha \, \laplacian{T} - u \,\frac{\partial T}{\partial z} = 0, 
}{eq:energy-1}
where $\alpha$ is the thermal diffusivity of the fluid.
The boundary condition usually with slip flow can be written in a general form:
\begin{gather}\label{eq:temp-jump}
T - \Tw + \lambda_T \, \vb{n} \vdot \grad{T} = 0
\qc (x,y) \in \Pcal
\end{gather}
where $\Tw(x,y)$ is the solid wall temperature, and $T(x,y)$ at positions on $\Pcal$ represents the fluid temperature at the wall, which can be different from $\Tw$ due to the temperature jump condition.

Despite the temperature jump condition associated with the slip flow configuration, as defined by eq.~\eqref{eq:temp-jump}, the thermal boundary condition envisaged in this study is the so-called {\sf H1} condition, according to the classification defined by \citet{shah1978laminar}. 
Recall that this condition means a peripherally (along $\Pcal$) uniform wall temperature and a longitudinally uniform wall heat flux. 
As a result, $\Tw$ is not known a priori, as it is the peripherally-averaged flux that is specified. The heat flux is written as
\begin{gather}\label{eq:wall-heat-flux}
\qw \,=\, k\,\vb{n} \vdot \grad{T} \qc (x,y) \in \Pcal
\end{gather}
and the peripherally-averaged quantities are given by:
\eqn{
\Twb \,=\, \frac{1}{\Pcal}\,\int_\Pcal \Tw\:\dd\Pcal 
\qc
\qwb \,=\, \frac{1}{\Pcal}\,\int_\Pcal \qw\:\dd\Pcal
}{}
which, due to the ${\sf H1}$ condition, $\qwb$ is a known constant and $\Twb = \Tw$, which is a function of $z$ alone.
This boundary condition configuration implies that the derivative $\partial T/\partial z$ is a constant \cite{shah1978laminar, BARLETTA199615}, hereafter denoted as $a$.
Also, it is easily verified that the constant $a$ not only coincides with $\partial T/ \partial z$, but also with $\dd \Twb/\dd z$ and $\dd \Tb/\dd z$, where $\Tb$ is the bulk temperature,
\eqn{
\pdv{T}{z} = \dv{\Twb}{z} = \dv{\Tb}{z} = a  \qc \Tb = \frac{1}{\Scal\, \um} \int\limits_\Scal T\, u \: \dd\Scal ,
}{eq:T-Props+BulkDef}
where $\um$ is the cross-section average velocity:
\begin{gather}\label{eq:um-def}
\um = \frac{1}{\Scal} \int\limits_\Scal u \, \dd \Scal\,
\end{gather}


Having defined the previous parameters, the Nusselt number is given by~\cite{shah1978laminar}:
\eqn{
\Nu \,=\, \frac{\hb\,\DhNom}{k} \,=\, \frac{\DhNom}{k} \, \frac{\qwb}{\Twb - \Tb}
}{eq:NusseltDef}
where $\hb$ is the cross-section-averaged convective heat transfer coefficient.
At this point, one should highlight that, a priori, the only known information about the duct geometry is that, nominally, its cross section is  circular with a  diameter $\DhNom$. For this reason the nominal diameter $\DhNom$ is used in the definition of the Nusselt number.
If the channel were perfectly smooth $\qwb$, $\Twb$  and $\Tb$ would match that for a perfect circular geometry. However, the roughness will have a tendency to require a larger temperature difference $\Twb - \Tb$ for the same average heat flux $\qwb$, due to an additional thermal resistance, such that Nusselt number values smaller than those obtained for smooth channels can be anticipated.


\subsection{Dimensionless formulation}

Dimensionless quantities and operators are defined as
\eqn{
(\xs, \ys) = \frac{(x,y)}{\rmax} \qc \nablas = \rmax \,\nabla \qc 
\us = \frac{u}{u_0} \nonumber \\
\Ts = k\,\frac{T - \Twb}{\qwb\, \rmax}  \qc
\lambdas = \frac{\lambda}{2\,\rmax} \qc
\lambdas_T = \frac{\lambda_T}{2\,\rmax} ,
}{4}
where $k$ is the thermal conductivity of the fluid and
\eqn{
u_0 = - \frac{2 \rmax^2}{\mu}\, \dv{p}{z} 
}{eq:u0-def} 
is a constant reference velocity such that
\eqn{
\frac{u_0}{\um} = f \, \Re = \sigma \qc 
}{6}
where $u_m$ is the mean flow velocity, $f$ is the Fanning friction factor and $\Re =  u_m\, \DhNom/\nu$ is the Reynolds number \cite{shah1978laminar}.
On account of \eqref{eq:u0-def} and \eqref{eq:T-Props+BulkDef}, the dimensionless temperature $\Ts$ defined by \eqref{4} possess the property
\eqn{
\dv{\Ts}{z} = 0 .
}{8}
On account of \eqref{4}, one can rewrite \eqref{eq:mom-1} as
\eqn{
{\nablas}^2 \us + \frac{1}{2} = 0 ,
}{10}
to be solved by employing the boundary condition
\eqn{
\us + 2\,\lambdas\,\vb{n} \vdot \nablas \us = 0 \qc (\xs,\ys) \in \Ps.
}{11}
The symbols $\Ps$ and $\Ss$ denote the dimensionless geometric objects obtained from $\Pcal$ and $\Scal$ with the coordinates scaled by the reference length $\rmax$. 
After solving \eqref{10} and \eqref{11}, one can evaluate the Poiseuille number $\sigma = f\, \Re$ by employing \eqref{eq:u0-def} and \eqref{6} through the formula
\eqn{
\sigma = f\,\Re = \left(\frac{1}{\Ss}\,\int_{\Ss} \us \, \dd \Ss \right)^{-1}.
}{13}
On account of \eqref{4}, the local energy balance equation \eqref{eq:energy-1} can be rewritten as
\eqn{
{\nablas}^2\,\Ts- \sigma\, \as\, \us  = 0 , 
}{eq:energy-DL}
where $\as$ is a dimensionless parameter,
\eqn{
\as \,=\, 
\frac{k \, \rmax \, \um}{\qwb \, \alpha}\, \frac{\partial T}{\partial z} .
}{15}
By employing Gauss' theorem together with the definitions given by \eqref{4} and \eqref{eq:u0-def}, an integration of \eqref{eq:energy-DL} over the dimensionless region $\Ss$ relates $\as$ directly to the dimensionless hydraulic diameter:
\eqn{
\as = \frac{\Ps}{\Ss}  =  \frac{2}{\Dhs}.
}{16}
where $\Dhs = \Dh/\rmax$.

The boundary conditions to be satisfied are, alternatively,
%
%
\eqn{
\Ts \,+\,  2\, \lambdas_T \, \vb{n} \vdot \nablas \Ts  \,=\, 0 \qc
(\xs,\ys) \in \Ps 
}{eq:BD-energy-DL}
%
%
%
After solving the previous equation toward the determination of $\us$ and $\Ts$, one can finally evaluate the Nusselt number $\Nu$,  by employing equations \eqref{eq:T-Props+BulkDef} to  eq.~\eqref{eq:NusseltDef}:
\eqn{
\Nu \,=\, \frac{2}{\Twbs - \Tbs},
}{eq:Nusselt-Calc}
where the dimensionless bulk temperature is calculated from
\begin{gather}
\Tbs \,=\, \frac{\sigma}{\Ss}  \int_{\Ss} \Ts \, \us \, \dd \Ss,
\end{gather} 
and, naturally, $\Twbs=0$ due to the considered {\sf H1} condition. %

\subsection{A perfectly smooth round microchannel\label{sec:nominal}}

Recalling the well-known results relative to an exactly circular cross-section may be useful for the forthcoming discussion of the results. In the absence of any roughness, the dimensionless velocity is given by modified Hagen-Poiseuille profile due to the presence of the slip condition:
\eqn{
\bar{u} = \frac{1}{8} \qty(1 - {\rs}^2 + 4\,\lambdas) ,
}{22}
with $\rs = \sqrt{{\xs}^2 + {\ys}^2}$.
Moreover,
\eqn{
\Ss = \pi \qc \bar\Pcal = 2\pi . 
}{23}
Thus, \eqref{13} and \eqref{16} yield
\eqn{
\sigma = f\, \Re = \frac{16}{8\, \lambdas +1} 
\qc \as = 2 .
}{24}
One can now employ \eqref{eq:energy-DL} and  \eqref{eq:BD-energy-DL} in a perfect circle to yield:
\eqn{
\Ts = -2 \, \lambdas_T  - \frac{\left({\rs}^2-1\right) \left(-16\, \lambdas +{\rs}^2-3\right)}{32 \lambdas +4} , 
}{} 
such that the Nusselt number leads to:
\begin{gather}
\Nu = \frac{48 (8 \,\lambdas +1)^2}{48 (8\,\lambdas +1)^2  \lambdas_T+128 \lambdas  (3\, \lambdas +1)+11} .
\end{gather}

\section{Numerical methodology}

\begin{figure}[h!]
\centering
\includegraphics[width=0.45\columnwidth]{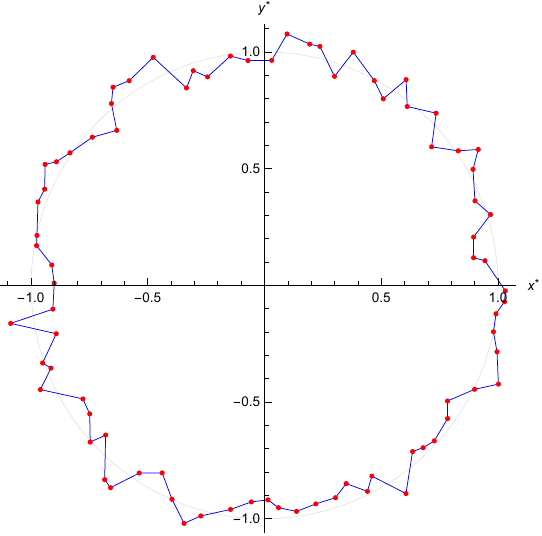}(a) \includegraphics[width=0.45\columnwidth]{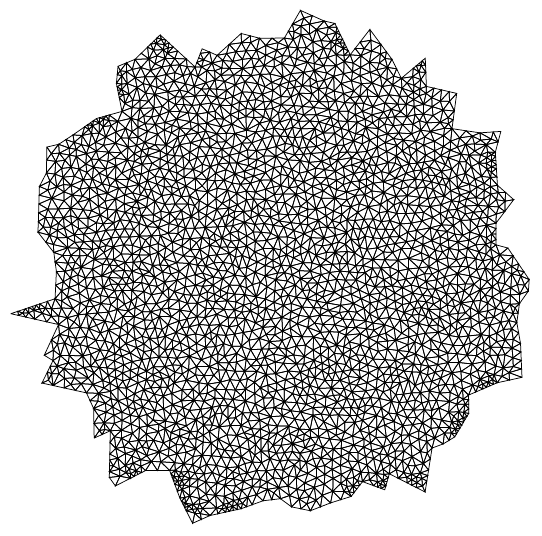}(b)
\caption{\label{fig1}\rm\small (a) Sample polygonal approximation of the actual microchannel boundary (blue) as compared to the smooth circular boundary (gray); (b) Unstructured mesh for the actual computational domain}
\end{figure}

The dimensionless boundary $\Ps$ of the microchannel cross-section $\Ss$  is defined by a polygon as displayed in  Fig.~\ref{fig1}. 
The points connected by the polygonal path are randomly generated with polar coordinates by assigning angles between $0$ and $2\pi$ and radii within the range $[0.9, 1.1]$. 
In Fig.~\ref{fig1}, the number of points used for generating the polygon is $40$. 
The computational domain is the dimensionless region bounded by the polygonal path. Uniform unstructured meshes are generated with triangular elements. The domain and mesh generation is achieved by using the software {\sl Mathematica 14} (\copyright{~}Wolfram Research Inc.). Furthermore, the weak formulation of the governing equations and their solution with the finite element method is managed by employing the built-in function {\bf NDSolve}. 
The accuracy of the numerical solution can be tested by inspecting the effect of increasing refinements of the mesh. Refining the mesh is possible by prescribing the parameter {\bf MaxCellMeasure} in the functions {\bf DiscretizeRegion}, {\bf ToElementMesh} and {\bf ToBoundaryMesh}, used for the discretization. 
An accuracy test is done with reference to the region displayed in Fig.~\ref{fig1} for the evaluation of $f\,\Re$ and $\Nu$ for $\lambdas=0.1$ and $\lambdas_T = 2\,\gamma \,\lambdas \big / ((\gamma +1)\, \Pr )$, where air properties at room temperature were employed for evaluating $\gamma$ and $\Pr$ (i.e.~$\gamma = 1.4$ and  $\Pr = 0.7$). The results are displayed in table~\ref{tab:convergence}.
\begin{table}[h!]
\centering
\begin{tabular}{|c|c|c|}
\hline
{\bf MaxCellMeasure} & $f\,\Re$ & $\Nu$ \\
\hline
 0.1 & 16.0738 & 4.36919 \\ \hline
 0.01 & 16.0702 & 4.36829 \\ \hline
 0.001 & 16.0684 & 4.36816 \\ \hline
 0.0001 & 16.0683 & 4.36816 \\ \hline
 0.00001 & 16.0683 & 4.36816 \\
\hline
\end{tabular}
\caption{\label{tab:convergence}\rm\small Accuracy test of the numerical solution}
\end{table}
As can be seen from these convergence results, by decreasing the value of {\bf MaxCellMeasure}, one attains accurate results with at least four significant figures for ${\bf MaxCellMeasure} = 10^{-3}$. In order to keep the computational time reasonably small,  for each case considered, all  forthcoming results are generated with this value.
%
%
%

\section{Results and discussion}

\begin{figure}[h!]
\centering
\includegraphics[width=0.47\columnwidth]{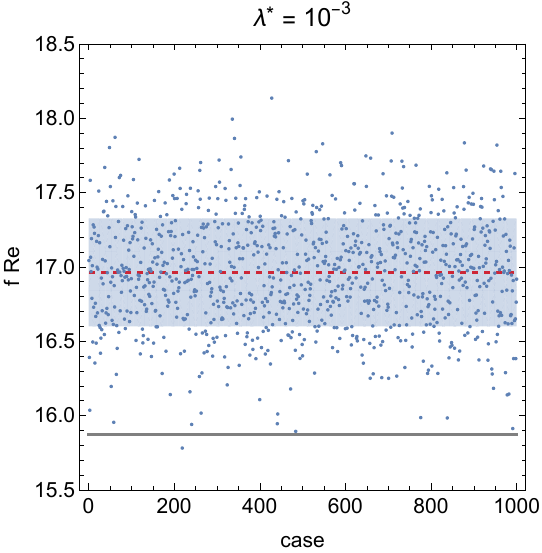}
\includegraphics[width=0.47\columnwidth]{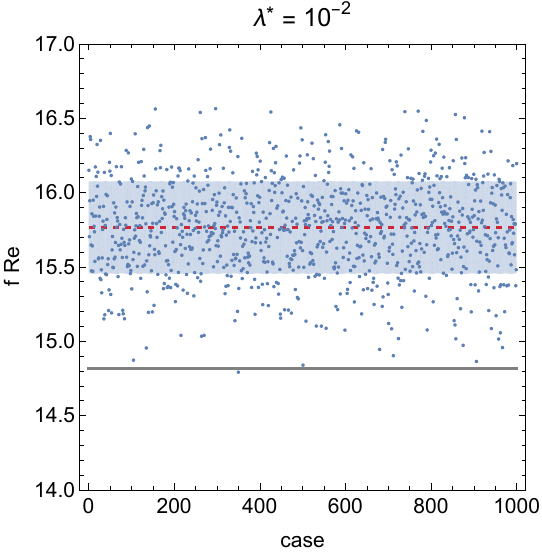}
\\
\includegraphics[width=0.47\columnwidth]{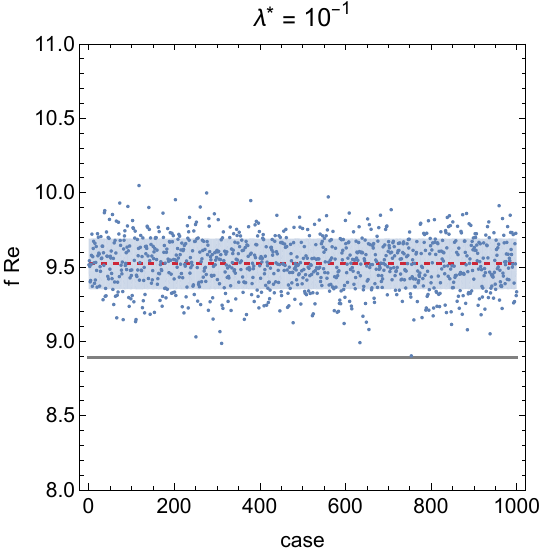}
\caption{\label{fig:Po}\rm\small Values of $\sigma = f\,\Re$ (blue dots) for the 1000 different geometries; the red dashed line indicates the mean value $\langle\sigma\rangle$ and the blue band denotes the range $\langle\sigma\rangle\pm\Delta\sigma$, where $\Delta\sigma$ is the standard deviation; the grey line represents the nominal value of $\sigma$, relative to the smooth circular microchannel.}
\end{figure}

\begin{figure}[h!]
\centering
\includegraphics[width=0.47\columnwidth]{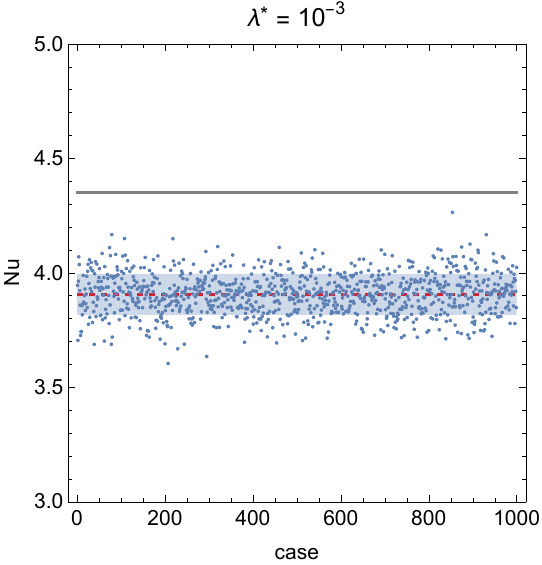}
\includegraphics[width=0.47\columnwidth]{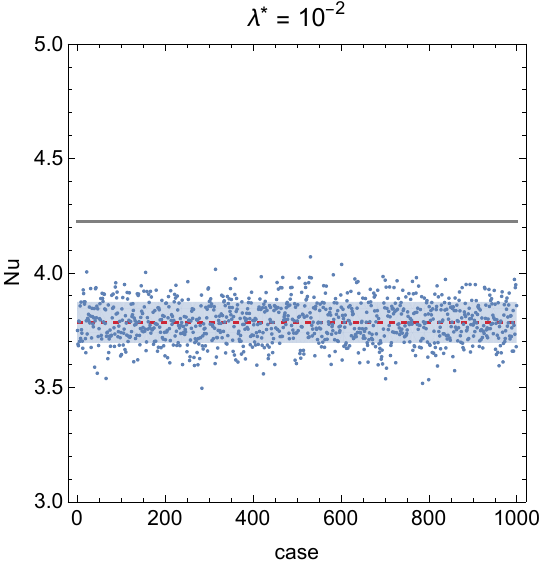}
\\
\includegraphics[width=0.47\columnwidth]{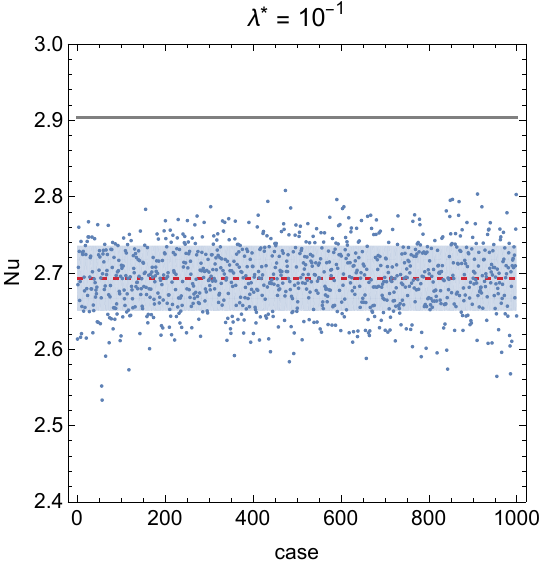}
\caption{\label{fig:Nu}\rm\small Values of $\sigma = f\,\Re$ (blue dots) for the 1000 different geometries; the red dashed line indicates the mean value $\langle\sigma\rangle$ and the blue band denotes the range $\langle\sigma\rangle\pm\Delta\sigma$, where $\Delta\sigma$ is the standard deviation; the grey line represents the nominal value of $\sigma$, relative to the smooth circular microchannel.}
\end{figure}

A statistical sample of $1000$ different microchannels is obtained by generating multiple unique polygonal paths and, hence, multiple computational regions where the values of $f\,\Re$ and $\Nu$, for different slip length cases  are evaluated, namely $\lambda = 10^{-3}$, $10^{-2}$ and $10^{-1}$. 
The values of $\lambdas_T$ were again calculated using $\lambdas_T = 2\,\gamma \,\lambdas \big / ((\gamma +1)\, \Pr )$ and considering air properties at room temperature.
Such values have been reported in Figs.~\ref{fig:Po} and \ref{fig:Nu} where, in abscissa, the labels of the 1000 different microchannels are given. 
Also, for complementing these figures, table~\ref{tab:results} present quantitative information calculated from the plotted data.
In these figures, the nominal values of $f\,\Re$ and $\Nu$ as calculated through the expressions in section~\ref{sec:nominal} are plotted for comparison as grey lines, whereas the actual mean values are identified as red dashed lines with a blue colored band showing the interval within the standard deviation tolerance. 
\begin{table}[h!]
\centering
\begin{tabular}{|c|c|c|c|c|c|}
\hline
$\lambdas$ & nominal &  mean & \%dif. & st.dev. & st.dev(\%)  \\
\hline
\multicolumn{5}{c}{$f\,\Re$} \\
\hline
0 	        & 16.000 & 17.133 & 7.08\% & 0.348 & 2.03\%\\  \hline
$10^{-3}$ & 15.873 & 16.962 & 6.86\% & 0.364 & 2.15\% \\  \hline
$10^{-2}$ & 14.815 & 15.764 & 6.41\%& 0.311 & 1.98\% \\ \hline
$10^{-1}$ & 8.8889 & 9.5195 & 7.09\% &  0.171 & 1.80\% \\ 
\hline
\multicolumn{5}{c}{$\Nu$} \\
\hline
0 	        & 4.3636 & 3.9108 & 10.4\% & 0.0926 & 2.37\% \\ \hline
$10^{-3}$ & 4.3508 & 3.9050 & 10.2\% & 0.0896 & 2.30\% \\ \hline
$10^{-2}$ & 4.2250 & 3.7832 & 10.5\% & 0.0897 & 2.37\% \\ \hline
$10^{-1}$ & 2.9037 & 2.6928 & 7.26\% & 0.0428 & 1.59\% \\ 
\hline
\end{tabular}
\caption{\label{tab:results}\rm\small Data for $f\,\Re$ and $\Nu$ and associated quantities for different values of $\lambdas$}
\end{table}
In table~\ref{tab:results}, besides reporting the mean values and standard deviation (in absolute value and \%) for the 1000 geometry cases, the calculated nominal values of $f\,\Re$ and $\Nu$ are presented as well, and the relative difference, calculated as $|f_{nominal} - f_{mean}|/f_{nominal}$, where $f$ is the quantity of interest, is also portrayed.
As one can observe, in all cases, it turns out that the values obtained analytically for the circular duct are more than one standard deviation away from the mean values. Such a result is both evident from Figs.~\ref{fig:Po} and \ref{fig:Nu} as well as from Table~\ref{tab:results}. As a consequence, one may infer a systematic of the wall roughness.
There is a shape sensitivity of the values of $f\,\Re$ which results into a net increase of $f\,\Re$ when a surface roughness of the microchannel wall is allowed within a range of $\pm10\%$ of the duct radius. Conversely, the sensitivity of the Nusselt number values results in a net decrease of $\Nu$ when the roughness of the microchannel wall is allowed within the same range.
While such a roughness is unrealistic for macroscale circular ducts, it becomes quite conceivable as the size decreases to the microscale. 
The increase in $f\,\Re$ can be explained due to the augmented hydraulic resistance induced by the wall roughness. A side effect is the lower velocity close to the walls caused by the meandering geometry of the perimetral path enclosing the microchannel cross-section. 
Such a side effect may also explain the less efficient heat transfer at the wall and, hence, the lower Nusselt number with respect to the smooth circular case. 

When looking at the effect of the slip length parameter $\lambdas$ (and the resulting value $\lambdas_T$), one notices that the standard behavior seen for smooth microchannels where an increasing $\lambdas$ value leads to smaller $f\,\Re$ values (due to smaller friction at the wall) and smaller $\Nu$ values (due to the increased thermal resistance introduced by the temperature jump parameter $\lambdas_T$), is seen for the cases with rough walls. However, the values of $f\,\Re$ are larger and the values of $\Nu$ are smaller, when compared to smooth channels, due to the roughness effect.

Although there is a clear trend where the roughness causes an increase in $f\,\Re$ and decrease in $\Nu$, as previously explained, a small number of exceptions is seen. For these cases, a particular configuration of the geometry points leads to a specific combination of flow area $\Ss$ and perimeter $\Ps$, which leads to $f\,\Re$ values that are smaller than the nominal value or $\Nu$ values that are larger than the nominal values. Nevertheless, this is far from the statistically most expected behavior.
%
%
%
%
As noted by \citet{pelevic2016heat}, the reason behind a predicted heat transfer enhancement or reduction remains unclear. In fact, as conjectured by \citet{CROCE2004601} and by \citet{pelevic2016heat}, the heat transfer rate is expected to be unaffected by surface roughness on the average. Such a conjecture is not confirmed by the present analysis.

\section{Conclusions}

Laminar forced convection in a circular microchannel has been investigated by considering the effect of slip-flow and and wall roughness.  The wall roughness has been assumed to affect the wall geometry in a random way within a range of $\pm 10\%$ of the duct radius. 
A statistical sample of one thousand different microchannels was generated as computational domains to be employed for the solution of the governing equations. In each instance of the statistical sample, cases with different slip-flow conditions were considered in order to evaluate the Nusselt number $Nu$ and the Poiseuille number $\sigma = f\,\Re$.
The comparison between the actual microchannel and the smooth circular duct has revealed a larger Poiseuille number and a smaller Nusselt number in the rough walled microchannel. 

This analysis can be further extended by considering other boundary conditions, and other nominal geometries for the microchannel. 
Also, a parametric analysis of the effect of varying the roughness parameter beyond 10\% of the duct geometry may as well be conducted.
Finally, a more extensive statistical analysis of randomly generated data may be carried out.




\bibliographystyle{elsarticle-num-names} 
\bibliography{biblio}

\end{document}